\documentclass[twocolumn,conference]{IEEEtran}
\usepackage[T1]{fontenc}
\usepackage[latin9]{inputenc}
\usepackage{float}
\usepackage{amsmath}
\usepackage{amsthm}
\usepackage{amssymb}
\usepackage{graphicx}

\makeatletter

\floatstyle{ruled}
\newfloat{algorithm}{tbp}{loa}
\providecommand{\algorithmname}{Algorithm}
\floatname{algorithm}{\protect\algorithmname}

\theoremstyle{definition}
\newtheorem{defn}{\protect\definitionname}
\theoremstyle{plain}
\newtheorem{thm}{\protect\theoremname}
\theoremstyle{plain}
\newtheorem{prop}{\protect\propositionname}

\usepackage[caption=false,font=footnotesize]{subfig}
\usepackage{algpseudocode}

\algdef{SE}{Begin}{End}{\textbf{begin}}{\textbf{end}}
\algnewcommand\algorithmicforeach{\textbf{for each}}
\algdef{S}[FOR]{ForEach}[1]{\algorithmicforeach\ #1\ \algorithmicdo}

\makeatother

\providecommand{\definitionname}{Definition}
\providecommand{\propositionname}{Proposition}
\providecommand{\theoremname}{Theorem}

\begin{document}
\title{Task Allocation and Mobile Base Station Deployment in Wireless Powered
Spatial Crowdsourcing }
\author{\IEEEauthorblockN{Yutao Jiao\IEEEauthorrefmark{1}, Ping Wang\IEEEauthorrefmark{2},
Dusit Niyato\IEEEauthorrefmark{1}, Jun Zhao\IEEEauthorrefmark{1},
Bin Lin\IEEEauthorrefmark{3} and Dong In Kim\IEEEauthorrefmark{4}}\IEEEauthorblockA{\IEEEauthorrefmark{1}School of Computer Science and Engineering,
Nanyang Technological University, Singapore}\IEEEauthorblockA{\IEEEauthorrefmark{2}Lassonde School of Engineering, York University,
Canada}\IEEEauthorblockA{\IEEEauthorrefmark{3}College of Information Science and Technology,
Dalian Maritime University, China}\IEEEauthorblockA{\IEEEauthorrefmark{4}School of Information and Communication Engineering,
Sungkyunkwan University, Korea}}
\IEEEoverridecommandlockouts
\IEEEpubid{\makebox[\columnwidth]{978-1-5386-8099-5/19/\$31.00 \copyright2019 IEEE \hfill} \hspace{\columnsep}\makebox[\columnwidth]{ }}
\maketitle
\IEEEpubidadjcol
\begin{abstract}
Wireless power transfer (WPT) is a promising technology to prolong
the lifetime of sensor and communication devices, i.e., workers, in
completing crowdsourcing tasks by providing continuous and cost-effective
energy supplies. In this paper, we propose a wireless powered spatial
crowdsourcing (SC) framework which consists of two mutual dependent
phases: task allocation phase and data crowdsourcing phase. In the
task allocation phase, we propose a Stackelberg game based mechanism
for the SC platform to efficiently allocate spatial tasks and wireless
charging power to each worker. In the data crowdsourcing phase, the
workers may have an incentive to misreport its real working location
to improve its own utility, which manipulates the SC platform. To
address this issue, we present a strategyproof deployment mechanism
for the SC platform to deploy its mobile base station. We apply the
Moulin's generalized median mechanism and analyze the worst-case performance
in maximizing the SC platform's utility. Finally, numerical experiments
reveal the effectiveness of the proposed framework in allocating tasks
and charging power to workers while avoiding the dishonest worker's
manipulation.
\end{abstract}


\section{Introduction }

Integrated with advanced sensing and communication techniques, mobile
devices can help complete diverse location-aware tasks, such as large-scale
data acquisition and analysis of real-time traffic monitoring or weather
measurements at different places. By focusing on the geospatial data,
a new paradigm called \emph{spatial crowdsourcing (SC)~}\cite{Kazemi2012}
has received increasing attention in the last few years~\cite{Zhao2016,Guo2018}.
Typically, there are three entities in the SC system, including an
online SC platform, requesters and workers. As a core component of
the SC system, the SC platform serves as a broker which allows requesters
to post tasks and recruits workers to complete them. Each employed
worker then stays at or travels to its target task area to collect
and transmit the requested data back. We hence study the interactions
between the SC platform and the workers.

Most existing work assumes that there is always a reliable communication
infrastructure and enough energy available for workers to complete
the data transmission. However, this may not be realistic especially
when workers have to perform tasks in remote areas without a wireless
base station deployed. Moreover, the workers can be battery-powered
wireless mobile devices. Fortunately, studies~\cite{Peng2010,Li2018}
in wireless powered sensor networks have illustrated the feasibility
of using wireless power transfer (WPT)~\cite{Bi2015} in sensing
data collection to prolong the lifetime of sensors. In view of this,
we consider a paradigm\emph{ }called\emph{ wireless powered spatial
crowdsourcing} where the SC platform deploys a mobile base station
(BS), e.g., robotics, drones or vehicles to assist the data collection.
Moreover, the mobile BS serves as the infrastructure for communication
and wireless power transfer. 

\begin{figure}[tbh]
\begin{centering}
\includegraphics[width=0.45\columnwidth]{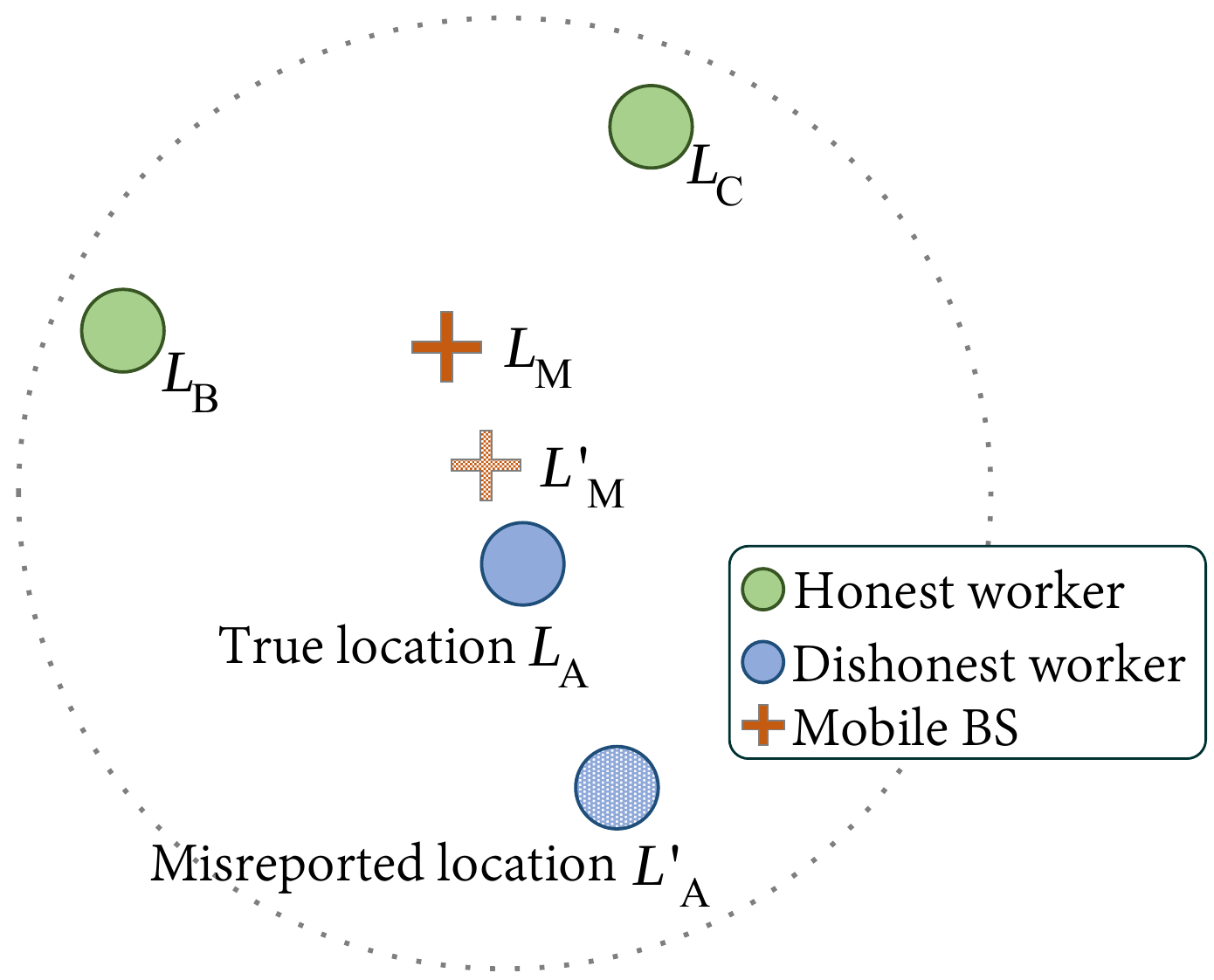}
\par\end{centering}
\caption{An example where a dishonest worker misreports its true location.\label{fig:illustrated-example}}
\end{figure}

To ensure a successful and stable operation of the crowdsourcing system,
designing an incentive mechanism that stimulates workers' participation
and efficiently allocates tasks is essential~\cite{Restuccia2016}.
A number of studies have proposed mechanisms satisfying various requirements,
such as profitability, strategyproofness, i.e., truthfulness, and
individual rationality~\cite{Yang2016}. Nevertheless, in wireless
powered crowdsourcing networks, the reward offered by the SC platform
to workers is the wireless power supply which is the major difference
with those existing mechanisms, the incentive of which is based on
monetary reward\footnote{The monetary reward can be tokens, virtual money, reputation, etc.}.
The difference introduces a few major issues for incentive mechanism
design in wireless powered crowdsourcing networks, and the following
questions have to be answered. First, what is the optimal total charging
power supply from the SC platform for maximizing its utility? The
SC platform can encourage workers to transmit sensed data at a higher
transmission rate, i.e., more collected data per unit time, but it
is at the cost of a higher power supply. Second, how to allocate the
tasks and charging power to workers which are spatially distributed
in the task area? The allocation is based on not only each worker's
sensing cost, but also the working location which clearly affects
the communication cost and transferred power. Note that the workers'
sensing cost and working location can be private information and unknown
to the SC platform. Lastly, how to deploy the mobile BS taking the
worker's strategic behavior into account? Since the worker's working
location is private, workers need to report their own locations before
the mobile BS chooses the best location to deploy. Under the assumption
of rationality, a worker may misreport its location to increase its
own utility while reducing the SC platform's utility. Figure~\ref{fig:illustrated-example}
shows a simple example. In the task area, there are one dishonest
worker at location $L_{\mathrm{A}}$ and two honest workers respectively
at $L_{\mathrm{B}}$ and $L_{\mathrm{C}}$. The SC platform should
place the mobile BS at $L_{\mathrm{M}}$ for optimal utility if all
workers report true locations $L_{\mathrm{A}},L_{\mathrm{B}}$ and
$L_{C}$. However, the dishonest worker has an incentive to report
a fake location $L'_{\mathrm{A}}$, so that according to the reported
locations $L'_{\mathrm{A}},L_{\mathrm{B}}$ and $L_{\mathrm{C}}$,
the mobile BS will be deployed at $L'_{\mathrm{M}}$. In this case,
the dishonest worker at $L_{\mathrm{A}}$ can be closer to the mobile
BS and then enjoy more transferred power from the mobile BS while
consuming less power to transmit its sensed data. Meanwhile, it inevitably
increases other workers' and SC platform's energy consumption and
damages their utilities finally. Most current studies on incentive
mechanisms for the crowdsourcing system have not addressed such issue
yet. 
In this paper, we propose a strategyproof and energy-efficient SC
framework which jointly solves the problems of task and wireless charging
power allocation as well as the truthful working location reporting.
In the framework, there are two phases: task allocation phase and
data crowdsourcing phase. In the task allocation phase, the SC platform
determines and announces a fixed total charging power supply. All
workers interested in participating have to choose and submit their
preferred transmission rate for the corresponding portion of the supplied
power. We use the Stackelberg game to model the interactions between
workers and the SC platform, in which the SC platform determine each
worker's transmission rate and allocated power. Next, in the data
crowdsourcing phase, the mobile BS requests for workers' working locations.
Based on the Moulin's generalization median rule~\cite{Moulin1980},
the mobile BS applies the median mechanism in determining its service
location to deploy. Then, the workers start to perform sensing tasks.
We analyze the worst-case performance of the median mechanism for
maximizing the SC platform's utility. 

The rest of the paper is organized as follows. In Section~\ref{sec:System-Model:-Spatial},
we describe the system model of wireless powered spatial crowdsourcing.
Section~\ref{sec:Optimal-charging-power} proposes the task and charging
power allocation mechanism. In Section~\ref{sec:Bayesian-location-mechanism},
we present the strategyproof mechanism for mobile BS deployment in
data crowdsourcing phase. In Section~\ref{sec:Experimental-and-simulation},
we provide the experimental results. Finally, we conclude the paper
in Section~\ref{sec:Conclusion}. 

\section{System Model: Wireless Powered Spatial Crowdsourcing Market\label{sec:System-Model:-Spatial}}

\begin{figure}[tbh]
\begin{centering}
\includegraphics[width=0.8\columnwidth]{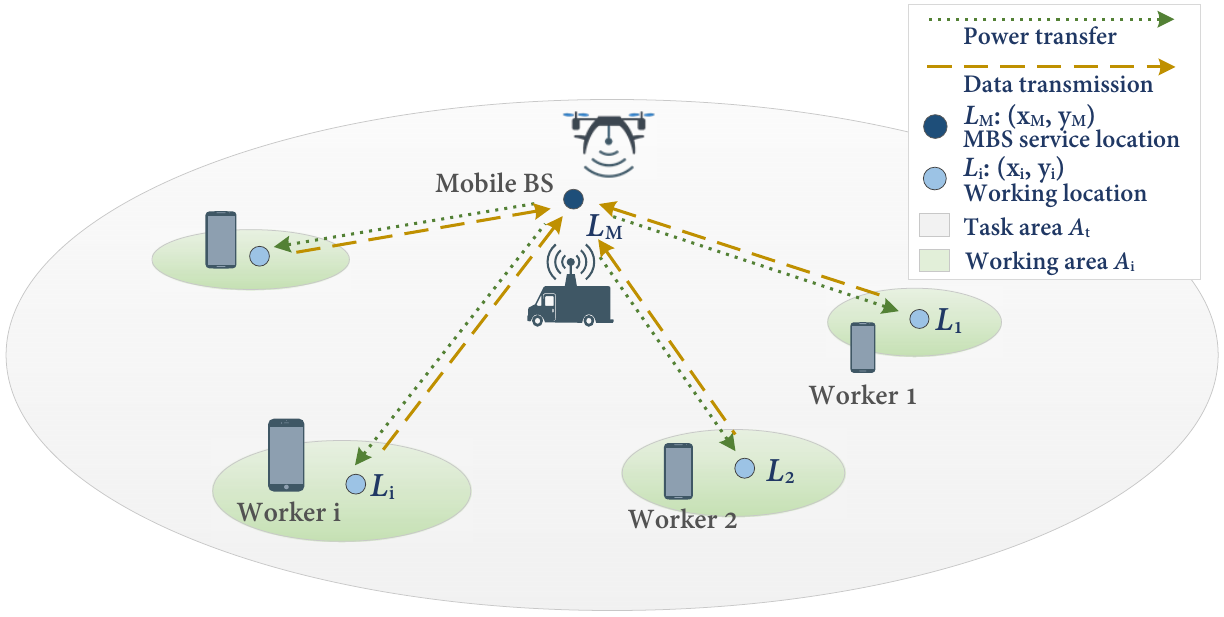}
\par\end{centering}
\caption{Data transmission and power transfer in the data crowdsourcing phase.\label{fig:Data-crowdsourcing-stage.}}
\end{figure}

The wireless powered spatial crowdsourcing system includes the requesters,
the SC platform residing in the cloud and the workers with mobile
sensing devices\footnote{The workers can be human, unmanned vehicles or robots.}.
Initially, the requesters publish spatial tasks with requirements,
such as the target task area, the task period and the sensed data
type. Then, the SC platform advertises the task information to \emph{registered
workers} on behalf of the requesters and collects the crowdsourced
data. As shown in Fig.~\ref{fig:Data-crowdsourcing-stage.}, we
denote by $\mathcal{N}=\{1,\ldots,N\}$ the set of workers and denote
by $\boldsymbol{A}_{t}$ the task area on a Cartesian coordinate plane.
The worker $i$'s working location $L_{i}$ is described by a 2-tuple,
i.e., $L_{i\in\mathcal{N}}=(x_{i},y_{i})$. We use $L_{\mathrm{M}}=(x_{\mathrm{M}},y_{\mathrm{M}})\in\boldsymbol{A}_{t}\subseteq\mathbb{R}^{2}$
to represent the deployed mobile BS's service location projected on
the XY-plane and use $h$ to denote its height. In the task area,
worker $i$ has its own working area $\boldsymbol{A}_{i}$ and its
working location $L_{i}$ falls in this area, i.e., $L_{i}\in\boldsymbol{A}_{i}\subseteq\boldsymbol{A}_{t}\subseteq\mathbb{R}^{2}$.

\subsection{Power cost model}

\subsubsection{Worker's power cost}

We consider an FDD system where sufficient channels are available
to ensure interference-free transmission. Moreover, we assume that
the communication channels are dominated by line-of-sight (LoS) links.
Given the mobile BS's service point $L_{\mathrm{M}}$, we can write
the worker $i$'s transmission rate according to Shannon's formula
as follows:
\begin{align}
r_{i} & =B\mathrm{\mathrm{\log_{2}}}\left(1+\frac{P_{i}^{\mathrm{t}}\delta}{\sigma^{2}d_{i}^{\alpha}}\right)=B\mathrm{\log_{2}}\left(1+\frac{P_{i}^{\mathrm{t}}g}{d_{i}^{\alpha}}\right)\label{eq:transmission_rate}
\end{align}
where $g=\frac{\delta}{\sigma^{2}}$ is the channel gain to noise
ratio (CNR), $\delta$ represents the corresponding channel power
gain at the reference distance of $1$ meter, $\sigma^{2}$ is the
noise power at the receiver mobile BS, $B$ is the channel bandwidth
in $\mathrm{hertz}$, $P_{i}^{\mathrm{t}}$ is worker $i$'s transmission
power, and $\alpha\geq2$ is the path-loss exponent. In addition,
we define 
\begin{align}
d_{i} & =d_{i}(L_{\mathrm{M}})=d((x_{i},y_{i}),(x_{\mathrm{M}},y_{\mathrm{M}}))\nonumber \\
 & =\sqrt{(x_{i}-x_{\mathrm{M}})^{2}+(y_{i}-y_{\mathrm{M}})^{2}+h^{2}}\label{eq:euclidean_distance}
\end{align}
as the euclidean distance between the worker $i$ and the mobile BS.
Note that $h$ is the height of the mobile BS. Hereby, we can derive
the worker $i$'s transmission power as
\begin{equation}
P_{i}^{\mathrm{t}}=\frac{(2^{\frac{r_{i}}{B}}-1)}{g}d_{i}^{\alpha}.\label{eq:transmission_power}
\end{equation}
Besides the power used to transmit data, for the worker $i$, we have
the power cost function of data sensing $P_{i}^{\mathrm{s}}=b_{i}r_{i}$
where $b_{i}$ is the energy cost per bit. Here, the power cost of
data sensing is linear to the sampling rate~\cite{Dieter2005}, i.e.,
the transmission rate. Therefore, the worker $i$'s total power cost
$P_{i}$ can be expressed as follows:
\begin{align}
P_{i} & =P_{i}^{\mathrm{t}}+P_{i}^{\mathrm{s}}=\frac{(2^{\frac{r_{i}}{B}}-1)}{g}d_{i}^{\alpha}+b_{i}r_{i}.\label{eq:worker_i_total_cost}
\end{align}

\subsubsection{Power cost of the mobile base station}

The mobile BS consumes energy mainly for WPT to workers. If the
charging power transferred to the worker $i$ is $P_{i}^{\mathrm{c}}$,
the mobile BS at the service location has to consume power $P_{i}^{\mathrm{c}'}$
derived as follows~\cite{Zhou2013}:
\begin{align}
P_{i}^{\mathrm{c}'} & =\frac{P_{i}^{\mathrm{c}}d_{i}^{\alpha}}{\eta\Gamma}=P_{i}^{\mathrm{c}}d_{i}^{\alpha}\kappa,\label{eq:charging_power}
\end{align}
where $\kappa=\frac{1}{\eta\Gamma}$, $0<\eta<1$ denotes the overall
receiver energy conversion efficiency, and $\Gamma$ denotes the combined
antenna gain of the transmit and receive antennas at the reference
distance of $1$ meter.

\subsection{Utility function in the wireless powered spatial crowdsourcing\textmd{\normalsize{}
}system\textmd{\normalsize{}}}

We evaluate the utility of the crowdsourced data based on the transmission
rate, which combines two common metrics, i.e., the data size and timeliness.
For example, the requesters may perform the data analysis and prediction
based on the real-time crowdsourced data. Higher data transmission
rate means that the requesters can process more data during a unit
time and yield more accurate prediction results. The quality of the
crowdsourced data is equivalent to the quality of the SC task completion.
We use a logarithmic function of the total data rate $R$ to characterize
the quality $q$ of the SC task completion as follows:
\begin{align}
q(R) & =a_{1}\log(1+a_{2}R)=a_{1}\log(1+a_{2}\sum_{i\in\mathcal{N}}r_{i}),\label{eq:sensing_quality}
\end{align}
where $R=\sum_{i\in\mathcal{N}}r_{i}$, $a_{1}$ and $a_{2}$ are
parameters. Taking the power cost of WPT~(\ref{eq:charging_power})
into consideration, the SC platform's utility function can be expressed
as
\begin{align}
u_{m} & =q(\tilde{R})-\sum_{i\in\mathcal{N}}P_{i}^{\mathrm{c}'}\nonumber \\
 & =a_{1}\log(1+a_{2}\sum_{i\in\mathcal{N}}r_{i})-\sum_{i\in\mathcal{N}}P_{i}^{\mathrm{c}}d_{i}^{\alpha}\kappa.\label{eq:SC_platform_utility_func}
\end{align}
Similarly, we obtain the worker $i$'s utility function as
\begin{align}
u_{i} & =P_{i}^{\mathrm{c}}-P_{i}=P_{i}^{\mathrm{c}}-\frac{(2^{\frac{r_{i}}{B}}-1)}{g}d_{i}^{\alpha}-b_{i}r_{i}.\label{eq:worker_i_utility_func}
\end{align}

\subsection{The procedure of wireless powered spatial crowdsourcing }

\subsubsection{Task allocation phase\label{subsec:Charging-Power-Allocation}}

First, the SC platform announces the task area $\boldsymbol{A}_{t}$
and a total charging power supply $P_{\mathrm{c}}$ ($P_{\mathrm{c}}=\sum_{j\in\mathcal{N}}P_{j}^{\mathrm{c}}$)
to assist workers in data crowdsourcing. The charging power $P_{i}^{\mathrm{c}}$
transferred to worker $i$ is proportional to the worker $i$'s contribution
(the data transmission rate) to the SC system, i.e., $P_{i}^{\mathrm{c}}=\frac{r_{i}}{R}P_{\mathrm{c}}=\frac{r_{i}}{\sum_{j\in\mathcal{N}}r_{j}}P_{\mathrm{c}}$.
Based on the sensing tasks and the other workers' responses, each
worker reports the preferred data rate $r_{i}$ to maximize its own
utility. We assume that the workers are \emph{risk-averse}, which
means that they choose to minimize the uncertainty and the possible
loss in the future. This concept can be found in the well-known prospect
theory~\cite{Kahneman2013}. A common example is that many people
prefer to deposit money at the bank for safe keeping and low return
instead of buying financial products with high risk of loss. As workers
have not set out to find the suitable working place and perform the
allocated task, they are exposed to the uncertainty of working location
$L_{i}$ and the mobile BS's service location $L_{\mathrm{M}}$ which
are only known in the next data crowdsourcing phase. Each worker plans
for the worst case where the mobile BS is deployed at the farthest
location from it. Since worker $i$ knows its own working area $\boldsymbol{A}_{i}$
and the task area $\boldsymbol{A}_{t}$, it can calculate the maximum
distance $D_{i}$ between $L_{i}$ and $L_{M}$, i.e., $D_{i}=\max_{L_{M}\in\boldsymbol{A}_{t},L_{i}\in\boldsymbol{A}_{i}}d_{i}$.
Note that the worker $i$'s utility function in~(\ref{eq:worker_i_utility_func})
is monotonically decreasing with $d_{i}$. When making the decision
on the transmission rate $r_{i}$, the worker $i$ sets $D_{i}$ as
the distance away from the mobile BS. Then, the worker cannot suffer
a loss due to the uncertain distance with the mobile BS in the data
crowdsourcing phase. We denote by $\mathbf{r}=(r_{1},r_{2},\ldots,r_{N})$
the workers' reported transmission rate vector. In addition, we use
$\mathbf{r}_{-i}=(r_{1},\ldots,r_{i-1},r_{i+1},\ldots,r_{N})$ to
denote the reported transmission rate vector for all workers except
the worker $i$. Hereby, the worker $i$'s utility function in the
task allocation phase can be expressed as
\begin{equation}
\bar{u}{}_{i}(r_{i},\mathbf{r}_{-i},P_{\mathrm{c}})=\frac{r_{i}}{\sum_{j\in\mathcal{N}}r_{j}}P_{\mathrm{c}}-\frac{(2^{\frac{r_{i}}{B}}-1)}{g}D_{i}^{\alpha}-b_{i}r_{i}.\label{eq:worker_i_utility_task_allocation_stage}
\end{equation}
The SC platform's utility in~(\ref{eq:SC_platform_utility_func})
is rewritten as
\begin{align}
\bar{u}{}_{m} & (P_{\mathrm{c}},\mathbf{r})=a_{1}\log(1+a_{2}\sum_{i\in\mathcal{N}}r_{i})-\sum_{i\in\mathcal{N}}\frac{r_{i}}{\sum_{j\in\mathcal{N}}r_{j}}P_{\mathrm{c}}D_{i}^{\alpha}\kappa.\label{eq:SC_platform_utility_task_allocation_phase}
\end{align}

\subsubsection{Data crowdsourcing phase\label{subsec:Data-crowdsoucing-phase}}

Once the total charging power supply $P_{c}$, each worker's allocated
charging power $P_{i}^{c}$ and transmission rate $r_{i}$ have been
determined in the task allocation phase, workers travel to the working
location and the SC platform sends out the mobile BS to serve the
workers. However, the mobile BS has to know each worker's working
location. Then, it can decide the service location $L_{\mathrm{M}}$
for maximizing the SC platform's utility. To make workers reveal
their private working location $L_{i}$, the mobile BS organizes
the following voting process on the spot. 
\begin{enumerate}
\item The mobile BS first broadcasts its deployment mechanism, i.e, the
mechanism or method to place the mobile BS according to the locations
reported by workers, to the task area. 
\item Once receiving the notification, each worker sends its working location
$L_{i}$ to the mobile BS.
\item Based on the collected locations and the deployment mechanism, the
service point location $L_{\mathrm{M}}$ is calculated for the mobile
BS to deploy.
\end{enumerate}
Let $\mathsf{M}$ denote the applied deployment mechanism which takes
the workers' reported working location vector $\mathbf{L}=(L_{1},\ldots,L_{i},\ldots,L_{\mathrm{M}})$
as input and outputs the mobile BS's service location $L_{\mathrm{M}}$,
i.e., $L_{\mathrm{M}}=\mathsf{M}(\mathbf{L})$. During the above voting
process, a worker $i$ may have the incentive to improve its own utility
in~(\ref{eq:worker_i_utility_func}) by misreporting its true working
location $L_{i}$. To make the location voting process robust and
implementable, our designed mobile BS deployment mechanism should
have the strategyproofness (truthfulness) property, which is defined
as follows: 
\begin{defn}
(Strategyproofness) Regardless of other workers' reported locations,
a worker $i$ cannot increase the utility by misreporting its working
location $L_{i}$. Formally, given a deployment mechanism $\mathsf{M}$
and a misreported location $L'_{i}$, we should have 
\begin{equation}
\hat{u}_{i}(\mathsf{M}((L_{i},\mathbf{L}_{-i})))\geq\hat{u}_{i}(\mathsf{M}((L'_{i},\mathbf{L}_{-i})))\:\forall L_{i}'\neq L_{i}
\end{equation}
where $\mathbf{L}_{-i}$ is the vector containing all workers' working
locations except the worker $i$'s.
\end{defn}

\section{Task and Wireless Transferred Power Allocation Mechanism\label{sec:Optimal-charging-power}}

We utilize the Stackelberg game~\cite{Fudenberg1991} to analyze
the model introduced in the task allocation phase (Section~\ref{subsec:Charging-Power-Allocation}).
Typically, there are two levels in the Stackelberg game. In the first
(upper) level, the SC platform acts as a leader which strategizes
and announces the total charging power supply $P_{c}$. In the second
(lower) level, each worker is a follower which determines the strategy,
i.e., the preferred transmission rate $r$, to maximize its own utility.
Mathematically, the SC platform chooses the strategy $P_{c}$ by solving
the following optimization problem:
\[
(\mathrm{P1})\;\max_{P_{c}\geq0}\bar{u}{}_{m}(P_{c},\mathbf{r}).
\]
Meanwhile, the worker $i$ makes the decision on its reported $r_{i}$
to solve the following problem:
\[
(\mathrm{P2})\;\max_{r_{i}\geq0}\bar{u}{}_{i}(r_{i},\mathbf{r}_{-i},P_{c}).
\]
The objective of the Stackelberg game is to find the \emph{Stackelberg
Equilibrium (SE)} points. We first introduce the concept of the SE
in our proposed model.
\begin{defn}
(Stackelberg Equilibrium) Let $\tilde{P}_{c}$ be a solution for Problem
$\mathrm{P1}$ and $\tilde{\mathbf{r}}$ be a solution for Problem
$\mathrm{P2}$ of the worker $i$. Then, the point $(\tilde{P}_{c},\tilde{\mathbf{r}})$
is a SE for the proposed Stackelberg game if it satisfies the following
conditions:
\begin{equation}
\bar{u}{}_{m}(\tilde{P}_{c},\tilde{\mathbf{r}})\geq\bar{u}{}_{m}(P_{c},\tilde{\mathbf{r}}),
\end{equation}
\begin{equation}
\bar{u}{}_{i}(\tilde{r}_{i},\tilde{\mathbf{r}}_{-i},\tilde{P}_{c})\geq\bar{u}{}_{i}(r_{i},\tilde{\mathbf{r}}_{-i},\tilde{P}_{c}),
\end{equation}
for any $(P_{c},\mathbf{r})$ with $P_{c}\geq0$ and $\mathbf{r}\succeq0$.
\end{defn}
In general, the first step to obtain the SE is to find the perfect
\emph{Nash Equilibrium (NE)~\cite{Fudenberg1991}} for the non-cooperative
transmission Rate Determination Game (RDG) in the second level. Then,
we can optimize the strategy of the SC platform in the first level.
Given a fixed $P_{c}$, the NE is defined as a set of strategies $\mathbf{r}^{\mathrm{ne}}=(r_{1}^{\mathrm{ne}},\ldots,r_{N}^{\mathrm{ne}})$
that no worker can improve its utility by unilaterally changing its
own strategy while other workers' strategies are kept unchanged. To
analyze the NE, we introduce the concept of \emph{concave game} and
the theorem about the \emph{existence} and \emph{uniqueness} of NE
in a concave game. 
\begin{defn}
(Concave game~\cite{Rosen1965})\emph{ \label{def:(Concave-game)}}A
game is called concave if each worker $i$ chooses a strategy $r_{i}$
to maximize utility $\bar{u}{}_{i}$, where $\bar{u}{}_{i}$ is concave
in $r_{i}$. 
\end{defn}
\begin{thm}
(\cite{Rosen1965}) \label{thm:NE-existance-uniqueness-Concave-games}Concave
games have (possibly multiple) Nash Equilibrium. Define $N\times N$
matrix function $\mathbf{H}$ in which $\mathbf{H}_{ij}=\frac{\partial^{2}\bar{u}{}_{i}}{\partial r_{i}\partial r_{j}}$,$i,j\in\mathcal{N}$.
Let \textup{$\mathbf{H}^{\mathrm{T}}$ denote the transpose of $\mathbf{H}$.}
If $\mathbf{H}+\mathbf{H}^{\mathrm{T}}$ is strictly negative definite,
then the Nash equilibrium is unique.
\end{thm}
Hereby, we calculate the first-order and second-order derivatives
of the worker $i$'s utility function $\bar{u}{}_{i}(r_{i},\mathbf{r}_{-i},P_{c})$
with respect to $r_{i}$ as follows: 
\begin{equation}
\frac{\partial\bar{u}{}_{i}}{\partial r_{i}}=\frac{P_{c}\sum_{k\in\mathcal{N}_{-i}}r_{k}}{(\sum_{j\in\mathcal{N}}r_{j})^{2}}-\frac{D_{i}^{\alpha}\ln2}{B}2^{\frac{r_{i}}{B}}-b_{i},\label{eq:first-derivative}
\end{equation}
\begin{equation}
\frac{\partial^{2}\bar{u}{}_{i}}{\partial r_{i}^{2}}=-\frac{2P_{c}\sum_{k\in\mathcal{N}_{-i}}r_{k}}{(\sum_{k\in\mathcal{N}}r_{k})^{3}}-\frac{D_{i}^{\alpha}\ln^{2}2}{B^{2}}2^{\frac{r_{i}}{B}}.\label{eq:second-derivative}
\end{equation}
Since $\frac{\partial^{2}\bar{u}{}_{i}}{\partial r_{i}^{2}}<0$, $\bar{u}{}_{i}(r_{i},\mathbf{r}_{-i},P_{c})$
is a strictly concave function with respect to $r_{i}$. Then, the
non-cooperative RDG is a concave game and the NE exists. Given any
$P_{c}>0$ and any strategy profile $\mathbf{r}_{-i}$ $(\sum_{j\in\mathcal{N}_{-i}}r_{j}>0)$,
the worker $i$'s best response strategy $\gamma_{i}$ exists and
is unique. To prove the uniqueness of the NE, we also calculate the
second-order mixed partial derivative of $\bar{u}{}_{i}$ for $i\in\mathcal{N}$
with respect to $r_{j\in\mathcal{N}_{-i}}$ as follows: 
\[
\frac{\partial^{2}\bar{u}{}_{i}}{\partial r_{j}^{2}}=\frac{2r_{i}}{(\sum_{j\in\mathcal{N}}r_{j})^{3}}P_{\mathrm{c}},\,\,\frac{\partial^{2}\bar{u}{}_{i}}{\partial r_{i}\partial r_{j}}=\frac{r_{i}-\sum_{k\in\mathcal{N}_{-i}}r_{k}}{(\sum_{k\in\mathcal{N}}r_{k})^{3}}P_{\mathrm{c}},
\]
where $\frac{\partial^{2}\bar{u}{}_{i}}{\partial r_{j}^{2}}\geq0$
and $\frac{\partial^{2}\bar{u}{}_{i}}{\partial r_{i}\partial r_{j}}\leq0$
if $r_{i}\leq\sum_{k\in\mathcal{N}_{-i}}r_{k},\,\forall i\in\mathcal{N}$.
Then, we have the specific expression of the matrix function $\mathbf{H}$
defined in Theorem~\ref{thm:NE-existance-uniqueness-Concave-games}.
Furthermore, the matrix function $\mathbf{H}+\mathbf{H}^{\mathrm{T}}$
can be decomposed into a sum of several $N\times N$ matrix functions:
$\mathbf{H}+\mathbf{H}^{\mathrm{T}}=\mathbf{U}+\mathbf{V}+\sum_{k\in\mathcal{N}}$$\mathbf{C}^{k}$,
where $\mathbf{U}_{ij}=\begin{cases}
0 & i\neq j\\
\frac{\partial^{2}\bar{u}{}_{i}}{\partial R_{i}^{2}} & i=j
\end{cases}$, $\mathbf{V}_{ij}=\sum_{k\in\mathcal{N}}\frac{\partial^{2}\bar{u}{}_{k}}{\partial r_{i}\partial r_{j}}$
and $\mathbf{C}_{ij}^{k}=$$\begin{cases}
0 & i=k\,\mathrm{or}\,j=k\\
-\frac{\partial^{2}\bar{u}{}_{k}}{\partial r_{i}\partial r_{j}} & otherwise.
\end{cases}$ Since $\frac{\partial^{2}\bar{u}{}_{i}}{\partial r_{i}^{2}}<0$ and
$\frac{\partial^{2}\bar{u}{}_{i}}{\partial r_{i}\partial r_{j}}\leq0$
if $r_{i}\leq\sum_{k\in\mathcal{N}_{-i}}r_{k},\,\forall i\in\mathcal{N}$,
we can find that $\mathbf{U}$ is strictly negative definite and $\mathbf{V},\,\sum_{k\in\mathcal{N}}\mathbf{C}^{k}$
are negative semi-definite. Thus, $\mathbf{H}+\mathbf{H}^{\mathrm{T}}$
is proved to be strictly negative definite which indicates the NE
in the RDG is unique. In other words, once the SC platform decides
a strategy $P_{\mathrm{c}}$, the workers' strategies, i.e., the transmission
rates, will be uniquely determined. We then can use the iterative
best response~\cite{Han2012} to find the SE\emph{ }point $\tilde{P}_{\mathrm{c}}$
in the first level, i.e., the optimal strategy of $P_{\mathrm{c}}$.

\section{Mobile BS Deployment Mechanism in Data Crowdsourcing Phase \label{sec:Bayesian-location-mechanism}}

Given the calculated SE points ($\tilde{P}_{c},\tilde{\mathbf{\mathbf{r}}}$)
from the task allocation phase, the specific problems for the SC platform
and workers in the data crowdsourcing phase are as follows,
\begin{align}
\max_{L_{M}\in\boldsymbol{A}_{t}}\hat{u}_{m}(L_{\mathrm{M}}) & =a_{1}\log(1+a_{2}\sum_{i\in\mathcal{N}}\tilde{r}_{i})-\nonumber \\
 & \,\,\,\,\,\,\sum_{i\in\mathcal{N}}\frac{\tilde{r}_{i}}{\sum_{j\in\mathcal{N}}\tilde{r}_{j}}\tilde{P_{c}}d_{i}^{\alpha}\kappa,\label{eq:SC_utility_data_crowdsourcing}
\end{align}
\begin{align}
\max_{L_{\mathrm{M}}\in\boldsymbol{A}_{t}}\hat{u}_{i}(L_{\mathrm{M}}) & =\frac{\tilde{r}_{i}}{\sum_{j\in\mathcal{N}}\tilde{r}_{j}}\tilde{P_{c}}-\frac{(2^{\frac{\tilde{r}_{i}}{B}}-1)}{g}d_{i}^{\alpha}-b_{i}\tilde{r}_{i},\label{eq:Worker_i_utility_data_crowdsourcing}
\end{align}
where $d_{i}=d_{i}(L_{\mathrm{M}})$ is defined in~(\ref{eq:euclidean_distance}).
To address the mobile BS's location problem introduced in Section~\ref{subsec:Data-crowdsoucing-phase},
we first introduce an important concept of \emph{$2$-dimensional
single-peaked preference }for the discussed problem\emph{.}

\begin{defn}
\emph{\label{def:2-dimensional-single-peaked}($2$-dimensional single-peaked
preference~\cite{Barbera1993})} Let $\mathbf{L}_{\mathrm{M}}$ be
the set of possible mobile BS's service locations output by the deployment
mechanism $\mathsf{M}$ on the XY-plane where $X$ and $Y$ are respectively
a one-dimensional axis. A worker $i$\textquoteright s preference
for the mobile BS's location is $2$-dimensional single-peaked with
respect to $(X,Y)$ if 1) there is a single most-preferred location
outcome $\tilde{L}_{i}^{\mathrm{M}}\in\mathbf{L}_{\mathrm{M}}$, and
2) for any two outcomes $L'_{\mathrm{M}},L''_{\mathrm{M}}\in\mathbf{L}_{\mathrm{M}}$,
$L'_{\mathrm{M}}\succeq_{i}L''_{\mathrm{M}}$ whenever $L''_{\mathrm{M}}<_{\rho}L'_{\mathrm{M}}<_{\rho}\tilde{L}_{i}^{\mathrm{M}}$
or $\tilde{L}_{i}^{\mathrm{M}}<_{\rho}L'_{\mathrm{M}}<_{\rho}L''_{\mathrm{M}}$
for $\forall\rho\in\{X,Y\}$, i.e., both $X$ and $Y$ axes. 
\end{defn}
In the above definition, $L'_{\mathrm{M}}\succeq_{i}L''_{\mathrm{M}}$
straightforwardly means $L'_{\mathrm{M}}$ is preferred by worker
$i$ to $L''_{\mathrm{M}}$. ``$<_{\rho}$'' is a strict ordering
by worker $i$ on the dimension $\rho$. An intuitive explanation
of this condition is that $L'_{\mathrm{M}}$ is preferred by worker
$i$ to $L''_{\mathrm{M}}$ as long as $L'_{\mathrm{M}}$ is nearer
to its most-preferred location $\tilde{L}_{i}^{\mathrm{M}}$ on each
dimension. 

\begin{prop}
\label{prop:2-dimensional-single-peaked-dcp}In the data crowdsourcing
phase, the worker's preference for the mobile BS's service location
is $2$-dimensional single-peaked.
\end{prop}
\begin{IEEEproof}
We first expand the worker $i$'s utility function given in~(\ref{eq:Worker_i_utility_data_crowdsourcing})
as $\hat{u}_{i}(L_{\mathrm{M}})=\hat{u}_{i}(x_{\mathrm{M}},y_{\mathrm{M}})=-\frac{(2^{\frac{\tilde{r}_{i}}{B}}-1)}{g}\left((x_{i}-x_{\mathrm{M}})^{2}+(y_{i}-y_{\mathrm{M}})^{2}+h^{2}\right)^{\frac{\alpha}{2}}+\frac{\tilde{r}_{i}}{\sum_{j\in\mathcal{N}}\tilde{r}_{j}}\tilde{P_{c}}-b_{i}\tilde{r}_{i}$.
It is clear that $\hat{u}{}_{i}$ is concave with respect to $(x_{\mathrm{M}},y_{\mathrm{M}})$
and there is a unique optimal solution $\tilde{L}_{i}^{\mathrm{M}}=(x_{i},y_{i})$
to maximizing the utility. In other words, the worker $i$'s most
preferred mobile BS's service location is its working location, i.e.,
$\tilde{L}_{i}^{\mathrm{M}}=L_{i}=(x_{i},y_{i})$, which satisfies
the first condition in Definition~\ref{def:2-dimensional-single-peaked}.
In the task area $\boldsymbol{A}_{t}$, we randomly choose two locations
$L'_{\mathrm{M}}=(x'_{\mathrm{M}},y'_{\mathrm{M}})$ and $L''_{\mathrm{M}}=(x''_{\mathrm{M}},y''_{\mathrm{M}})\in\boldsymbol{A}_{t}$.
Note that the concavity of $\hat{u}{}_{i}$ guarantees the concavity
on one dimension if fixing the variable on the other dimension. $L''_{\mathrm{M}}<_{X}L'{}_{\mathrm{M}}<_{X}\tilde{L}_{i}^{\mathrm{M}}$
implies that $\hat{u}{}_{i}((x''_{\mathrm{M}},y))<\hat{u}{}_{i}((x'_{\mathrm{M}},y))<\hat{u}{}_{i}((x_{i},y))$
for any $y$ on $Y$ axis and then$\left|x'_{\mathrm{M}}-x_{i}\right|<\left|x''_{\mathrm{M}}-x_{i}\right|$.
We can have the similar implication from $L''_{\mathrm{M}}<_{Y}L'{}_{\mathrm{M}}<_{Y}\tilde{L}_{i}^{M}$.
If $L''_{\mathrm{M}}<_{X}L'{}_{\mathrm{M}}<_{X}\tilde{L}_{i}^{\mathrm{M}}$
and $L''_{\mathrm{M}}<_{Y}L'_{\mathrm{M}}<_{Y}\tilde{L}_{i}^{M}$
are both satisfied, we can have $(x_{i}-x'_{\mathrm{M}})^{2}+(y_{i}-y'_{\mathrm{M}})^{2}<(x_{i}-x''_{\mathrm{M}})^{2}+(y_{i}-y''_{\mathrm{M}})^{2}$
and thus $\hat{u}_{i}(L''_{\mathrm{M}})=\hat{u}{}_{i}((x''_{\mathrm{M}},y''_{\mathrm{M}}))<\hat{u}_{i}(L'_{\mathrm{M}})=\hat{u}{}_{i}((x'_{\mathrm{M}},y'_{\mathrm{M}}))$.
Therefore, the worker $i$ prefers $L'_{\mathrm{M}}$ to $L''_{\mathrm{M}}$,
i.e., $L'_{\mathrm{M}}\succeq_{i}L''_{\mathrm{M}}$, which proves
the second condition in Definition~\ref{def:2-dimensional-single-peaked}
and completes the proof.
\end{IEEEproof}
\begin{thm}
\label{thm:(Moulin's-one-dimensional-genera)}(Moulin's one-dimensional
generalized median mechanism~\cite{Moulin1980}) A mechanism $\mathsf{M}$
for single-peaked preferences in a one-dimensional space is strategyproof
and anonymous if and only if there exist $N+1$ constants $\tau_{1},\ldots,\tau_{N+1}\in\mathbb{R}\cup(-\infty,+\infty)$
such that: 
\begin{equation}
\mathsf{M}(\mathbf{L}^{\mathrm{M}})=\mathrm{median}(\tilde{L}_{1}^{\mathrm{M}},\ldots,\tilde{L}_{N}^{\mathrm{M}},\tau_{1},\ldots,\tau_{N+1})
\end{equation}
\emph{where }\textup{\emph{$\mathbf{L}^{\mathrm{M}}=\{\tilde{L}_{1}^{\mathrm{M}},\ldots,\tilde{L}_{N}^{\mathrm{M}}\}$
is the set of workers' most-preferred mobile BS's locations and}}\emph{
}\textup{\emph{$\mathrm{median}$ is the median function. }}An outcome
rule $\mathsf{M}$ is anonymous, if for any permutation $\boldsymbol{T}'$
of $\boldsymbol{T}$, we have $\mathsf{M}(\boldsymbol{T}')=\mathsf{M}(\boldsymbol{T})$
for all $\boldsymbol{T}$.
\end{thm}
\begin{thm}
\label{thm:Multi-dimensional-generalized-median}(Multi-dimensional
generalized median mechanism~\cite{Barbera1993}) A mechanism for
multi-dimensional single-peaked preferences in a multi-dimensional
space is strategyproof and anonymous if and only if it is an $m$-dimensional
generalized median mechanism, which straightforwardly applies the
one-dimensional generalized median mechanism on each of the $m$ dimensions.
\end{thm}
\begin{algorithm}[tbh]
\scriptsize
\begin{algorithmic}[1]
\Require{Workers' reported locations $\mathbf{L}=(L_1,\dots, L_i, \dots,L_N)$.}
\Ensure{Mobile BS's service point location $L_M=(x_M,y_M)$.}
\Begin
	\State{Respectively sort the workers' locations on x-axis $\mathbf{x}=(x_1,\dots, x_N)$ and y-axis $\mathbf{y}=(y_1, \dots,y_N)$ in ascending order.}
	\If{$N$ is odd}
		\State{$x_M \gets x_{\frac{N+1}{2}}, y_M \gets y_{\frac{N+1}{2}}$}
	\Else
		\State{$x_M \gets \frac{x_{\frac{N}{2}}+x_{\frac{N}{2}+1}}{2}, y_M \gets \frac{y_{\frac{N}{2}}+y_{\frac{N}{2}+1}}{2}$}
	\EndIf 
\End
\end{algorithmic} 

\caption{MED mechanism\label{alg:1-median-mechanism}}
\end{algorithm}
A straightforward mechanism is the median mechanism~\cite{Moulin1980,Barbera1993},
as shown in Algorithm~\ref{alg:1-median-mechanism}. We simply name
it as MED mechanism $\mathsf{M}_{\mathrm{MED}}$. This algorithm directly
computes the median of workers' reported locations as the mobile BS's
service point location. Apparently, it is a special case of the multi-dimensional
generalized median mechanism, so it is strategyproof. We next analyze
its performance by comparing it with the optimal mechanism $\mathsf{M}_{\mathrm{OPT}}$
which achieves the maximum utility of the SC platform without considering
incentive constraints. Let $\tilde{r}_{\max},\tilde{r}_{\min}$
respectively denote the maximum and the minimum transmission rate
among workers, i.e., $\tilde{r}_{\max}=\max(\tilde{\mathbf{r}}),\tilde{r}_{\min}=\min(\tilde{\mathbf{r}})$
. 
\begin{prop}
\label{prop:The-approximation-ratio}The worst-case performance of
the MED mechanism $\mathsf{M}_{\mathrm{MED}}$ for maximizing the
SC platform's utility can guarantee\textup{ $\hat{u}{}_{m}(\mathsf{M}_{\mathrm{MED}}(\boldsymbol{L}))\geq\varphi-2^{\frac{\alpha}{2}}N^{\frac{\alpha}{2}-1}\frac{\tilde{r}_{\max}}{\tilde{r}_{\min}}(\varphi-\hat{u}{}_{m}(\mathsf{M}_{\mathrm{OPT}}(\boldsymbol{L})))$,
where $\varphi=a_{1}\log(1+a_{2}\sum_{i\in\mathcal{N}}\tilde{r}_{i})$.}
\end{prop}
\begin{IEEEproof}
We expand the SC platform's utility function in~(\ref{eq:SC_utility_data_crowdsourcing})
as follows:
\begin{align}
 & \hat{u}{}_{m}((x_{\mathrm{M}},y_{\mathrm{M}}))=\varphi-\nonumber \\
 & \frac{\tilde{P}_{c}\kappa}{\sum_{j\in\mathcal{N}}\tilde{r}_{j}}\sum_{i\in\mathcal{N}}\left(\tilde{r}_{i}^{\frac{2}{\alpha}}(x_{i}-x_{M})^{2}+(y_{i}-y_{M})^{2}+h^{2}\right)^{\frac{\alpha}{2}},\label{eq:SC-utility-rewitten-DCP}
\end{align}
where $\varphi=a_{1}\log(1+a_{2}\sum_{i\in\mathcal{N}}\tilde{r}_{i})$.
Let $x_{\mathrm{med}},\overline{x}$ and $y_{\mathrm{med}},\overline{y}$
respectively denote the median and mean of $\mathbf{x}=(x_{1},\ldots,x_{N})$
and $\mathbf{y}=(y_{1},\ldots,y_{N})$. Also, we use $(x_{\mathrm{opt}},y_{\mathrm{opt}})$
to denote the optimal solution to maximizing the expression in (\ref{eq:SC-utility-rewitten-DCP}),
i.e., $\mathsf{M}_{\mathrm{OPT}}(\boldsymbol{L})=(x_{\mathrm{opt}},y_{\mathrm{opt}})$.
We also note that the optimal solution to minimizing the $\sum_{i\in\mathcal{N}}\tilde{r}_{i}^{\frac{2}{\alpha}}\left((x_{i}-x_{\mathrm{M}})^{2}+(y_{i}-y_{\mathrm{M}})^{2}+h^{2}\right)$
is $(x^{*},y^{*})$ where $x^{*}=\frac{\sum_{i\in\mathcal{N}}\tilde{r}_{i}^{\frac{2}{\alpha}}x_{i}}{\sum_{i\in\mathcal{N}}\tilde{r}_{i}^{\frac{2}{\alpha}}}$
and $y^{*}=\frac{\sum_{i\in\mathcal{N}}\tilde{r}_{i}^{\frac{2}{\alpha}}y_{i}}{\sum_{i\in\mathcal{N}}\tilde{r}_{i}^{\frac{2}{\alpha}}}$.
As $\tilde{r}_{\min}\leq\tilde{r}_{i}$, we have
\begin{align}
 & \tilde{r}_{\min}^{\frac{2}{\alpha}}\sum_{i\in\mathcal{N}}\left((x_{i}-\overline{x})^{2}+(y_{i}-\overline{y})^{2}+h^{2}\right)\nonumber \\
 & \leq\sum_{i\in\mathcal{N}}\tilde{r}_{i}^{\frac{2}{\alpha}}\left((x_{i}-x^{*})^{2}+(y_{i}-y^{*})^{2}+h^{2}\right).
\end{align}
According to~\cite[Theorem 4.3]{Feldman2013}, we have $\sum_{i\in\mathcal{N}}(x_{i}-x_{\mathrm{med}})^{2}\leq2\sum_{i\in\mathcal{N}}(x_{i}-\overline{x})^{2}$
and $\sum_{i\in\mathcal{N}}(y_{i}-y_{\mathrm{med}})^{2}\leq2\sum_{i\in\mathcal{N}}(y_{i}-\overline{y})^{2}$.
Then, it is not hard to verify that 
\begin{align}
 & \tilde{r}_{\min}^{\frac{2}{\alpha}}\sum_{i\in\mathcal{N}}\left((x_{i}-x_{\mathrm{med}})^{2}+(y_{i}-y_{\mathrm{med}})^{2}+h^{2}\right)\nonumber \\
 & \leq2\tilde{r}_{\min}^{\frac{2}{\alpha}}\sum_{i\in\mathcal{N}}\left((x_{i}-\overline{x})^{2}+(y_{i}-\overline{y})^{2}+h^{2}\right),\\
 & \tilde{r}_{\min}\left(\sum_{i\in\mathcal{N}}\left((x_{i}-x_{\mathrm{med}})^{2}+(y_{i}-y_{\mathrm{med}})^{2}+h^{2}\right)\right)^{\frac{\alpha}{2}}\nonumber \\
 & \leq2^{\frac{\alpha}{2}}\tilde{r}_{\min}\left(\sum_{i\in\mathcal{N}}\left((x_{i}-\overline{x})^{2}+(y_{i}-\overline{y})^{2}+h^{2}\right)\right)^{\frac{\alpha}{2}}\nonumber \\
 & \leq2^{\frac{\alpha}{2}}\tilde{r}_{\min}\left(\sum_{i\in\mathcal{N}}\left((x_{i}-x^{*})^{2}+(y_{i}-y^{*})^{2}+h^{2}\right)\right)^{\frac{\alpha}{2}}\nonumber \\
 & \leq2^{\frac{\alpha}{2}}\left(\sum_{i\in\mathcal{N}}\tilde{r}_{i}^{\frac{2}{\alpha}}\left((x_{i}-x^{*})^{2}+(y_{i}-y^{*})^{2}+h^{2}\right)\right)^{\frac{\alpha}{2}}.\label{eq:inquality-med-mean}
\end{align}
Since $\alpha\geq2$, we can prove that
\begin{align}
 & \tilde{r}_{\min}\sum_{i\in\mathcal{N}}\left((x_{i}-x_{\mathrm{med}})^{2}+(y_{i}-y_{\mathrm{med}})^{2}+h^{2}\right)^{\frac{\alpha}{2}}\nonumber \\
 & \leq\tilde{r}_{\min}\left(\sum_{i\in\mathcal{N}}\left((x_{i}-x_{\mathrm{med}})^{2}+(y_{i}-y_{\mathrm{med}})^{2}+h^{2}\right)\right)^{\frac{\alpha}{2}}.
\end{align}
Hence, based on Theorem $1$ in~\cite{Jameson2014} and the facts
that $\tilde{r}_{i}\leq\tilde{r}_{\max}$ and $\frac{\alpha}{2}\geq1$,
we can obtain
\begin{align}
 & 2^{\frac{\alpha}{2}}\left(\sum_{i\in\mathcal{N}}\tilde{r}_{i}^{\frac{2}{\alpha}}\left((x_{i}-x^{*})^{2}+(y_{i}-y^{*})^{2}+h^{2}\right)\right)^{\frac{\alpha}{2}}\nonumber \\
 & \leq2^{\frac{\alpha}{2}}\left(\sum_{i\in\mathcal{N}}\tilde{r}_{i}^{\frac{2}{\alpha}}\left((x_{i}-x_{\mathrm{opt}})^{2}+(y_{i}-y_{\mathrm{opt}})^{2}+h^{2}\right)\right)^{\frac{\alpha}{2}}\nonumber \\
 & \leq2^{\frac{\alpha}{2}}\tilde{r}_{\max}\left(\sum_{i\in\mathcal{N}}\left((x_{i}-x_{\mathrm{opt}})^{2}+(y_{i}-y_{\mathrm{opt}})^{2}+h^{2}\right)\right)^{\frac{\alpha}{2}}\nonumber \\
 & \leq2^{\frac{\alpha}{2}}N^{\frac{\alpha}{2}-1}\tilde{r}_{\max}\sum_{i\in\mathcal{N}}\left((x_{i}-x_{\mathrm{opt}})^{2}+(y_{i}-y_{\mathrm{opt}})^{2}+h^{2}\right)^{\frac{\alpha}{2}}.
\end{align}
Combining the above inequalities, we have
\begin{align}
 & \varphi-\frac{\tilde{P}_{c}\kappa}{\sum_{j\in\mathcal{N}}\tilde{r}_{j}}\sum_{i\in\mathcal{N}}\left((x_{i}-x_{\mathrm{med}})^{2}+(y_{i}-y_{\mathrm{med}})^{2}+h^{2}\right)^{\frac{\alpha}{2}}\nonumber \\
 & \geq\varphi-\left(\frac{\tilde{P}_{c}\kappa}{\sum_{j\in\mathcal{N}}\tilde{r}_{j}}2^{\frac{\alpha}{2}}N^{\frac{\alpha}{2}-1}\frac{\tilde{r}_{\max}}{\tilde{r}_{\min}}\right.\nonumber \\
 & \,\,\,\,\,\,\,\,\,\,\,\,\,\,\,\,\,\,\,\,\left.\sum_{i\in\mathcal{N}}\left((x_{i}-x_{\mathrm{opt}})^{2}+(y_{i}-y_{\mathrm{opt}})^{2}+h^{2}\right)^{\frac{\alpha}{2}}\right).
\end{align}
Finally, we can conclude that $\hat{u}{}_{m}(\mathsf{M}_{\mathrm{MED}}(\boldsymbol{L}))\geq\varphi-2^{\frac{\alpha}{2}}N^{\frac{\alpha}{2}-1}\frac{\tilde{r}_{\max}}{\tilde{r}_{\min}}(\varphi-\hat{u}{}_{m}(\mathsf{M}_{\mathrm{OPT}}(\boldsymbol{L})))$.
\end{IEEEproof}

\section{Simulation results and discussions\label{sec:Experimental-and-simulation} }

In this section, we conduct simulations based on realistic data to
evaluate the performance of our proposed framework and strategyproof
deployment mechanisms. Unless otherwise stated, the simulation configuration
is set as follows. We consider a $[0,50]\times[0,50]$ square-meter
area as the SC task area $\mathrm{\boldsymbol{A}}_{t}$. The number
of registered workers is set as $N=50$. We generate worker $i$'s
working location from the bivariate uniform distribution, i.e., $\mathcal{P}^{\mathrm{u}}=\begin{cases}
1 & (x,y)\in\boldsymbol{A}=[25-\beta_{i},25+\beta_{i}]^{2}\\
0 & otherwise.
\end{cases}$, where $\beta_{i}\in\mathbb{R}$ is uniformly distributed in $[0,25]$.
Hereby, the maximum distance $D_{i}$ can naturally be calculated.
We set the height of the mobile BS $h=5\,\mathrm{m}$, the channel
gain to noise ratio $g=90\,\mathrm{dB}$, the bandwidth of each subchannel
$B=60\,\mathrm{MHz}$, the data utility parameters $a_{1}=1000$ and
$a_{2}=200$, the energy conversion efficiency $\eta=0.5$, the antenna
gain $\Gamma=-30\,\mathrm{dB}$, and the path-loss exponent $\alpha=2$.
The sensing energy cost per bit $b_{i}$ is generated from the uniform
distribution on $[10^{-4},1.5\times10^{-4}]$. Each measurement is
averaged over $100$ instances. 
\begin{figure}[tbh]
\begin{centering}
\includegraphics[width=0.65\columnwidth]{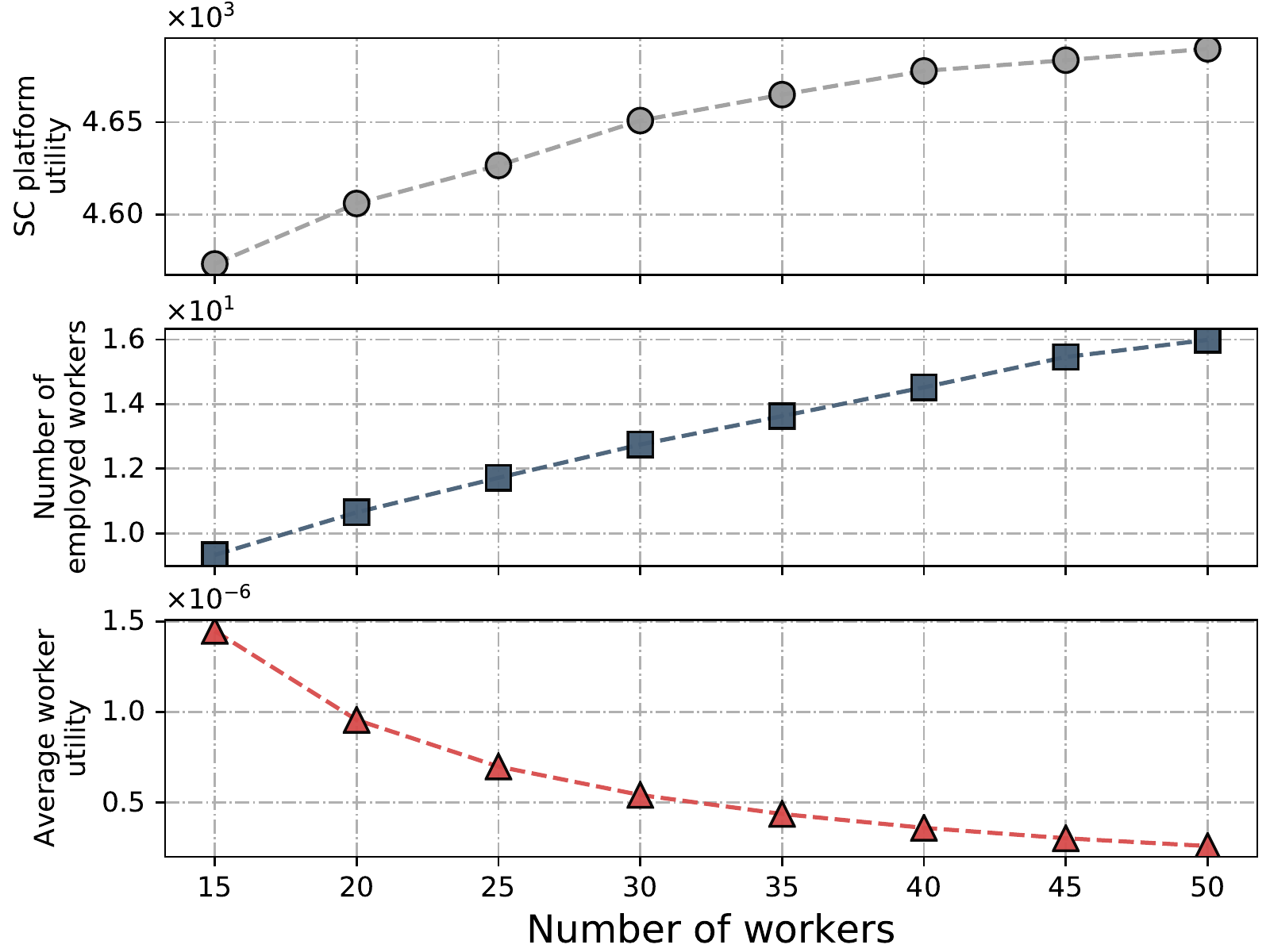}
\par\end{centering}
\caption{Impact of the number of workers on the SC platform's utility (top),
the number of employed workers (middle) and the average worker's utility
(bottom) in the task allocation phase.\label{fig:Impact_of_Ne}}
\end{figure}
\begin{figure}[tbh]
\begin{centering}
\includegraphics[width=0.65\columnwidth]{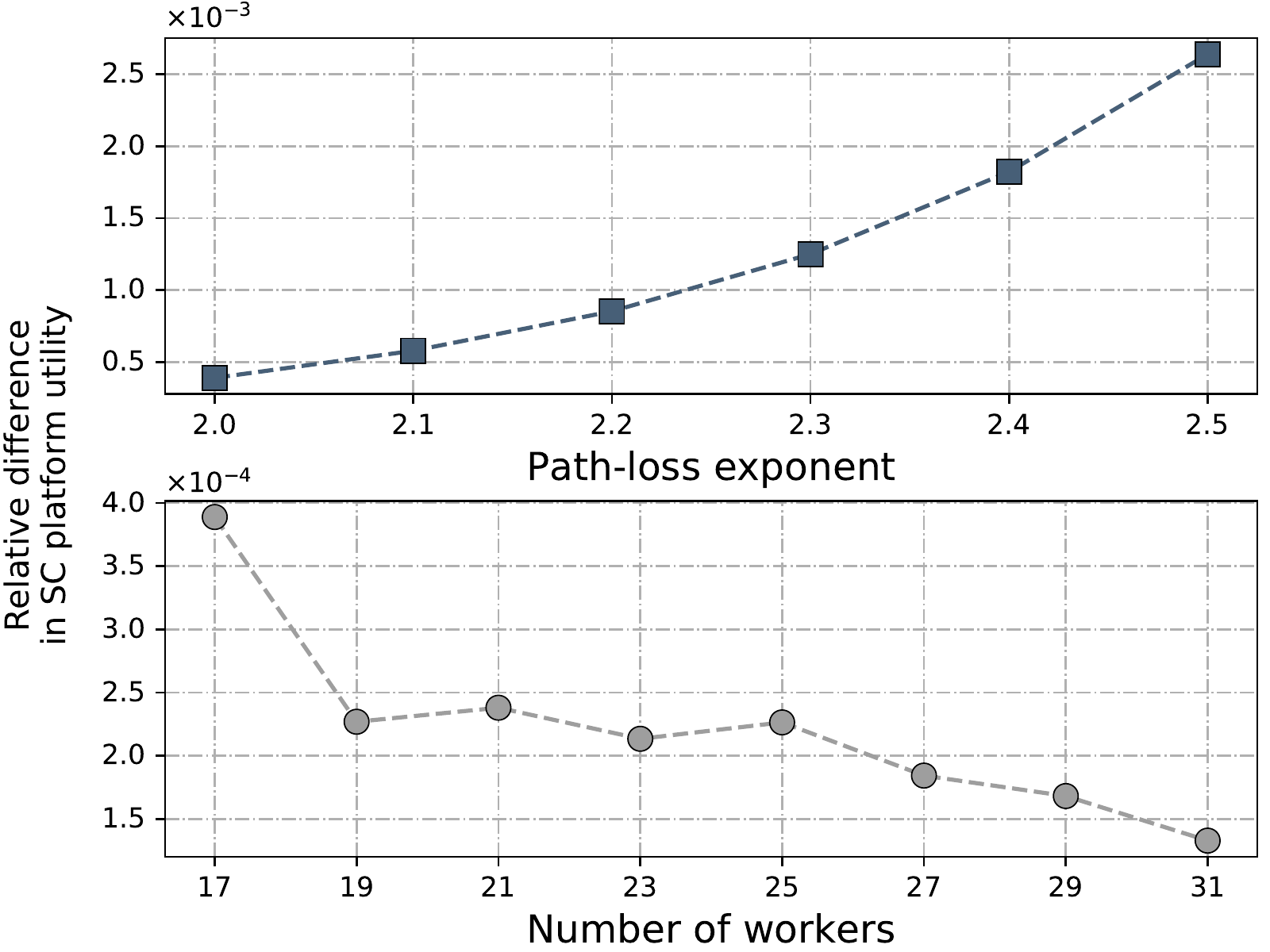}
\par\end{centering}
\caption{The SC platform utility achieved per different mechanism with varied
number of employed workers and varied path-loss exponent in the data
crowdsourcing phase.\label{fig:SC_utility_specialcase}}
\end{figure}
Figure~\ref{fig:Impact_of_Ne} demonstrates the impact of the number
of workers on the SC platform's utility, the average worker's utility
and the number of employed workers in the task allocation phase. When
the number of registered workers increases, the SC platform's utility
and the number of employed workers gradually increase but with a diminishing
return. These results reflect that when more workers are employed,
the SC platform has to consume more charging power for the same marginal
utility. By contrast, the average worker's utility decreases with
the increase of registered workers because of the more fierce competition
among workers. Figure~\ref{fig:SC_utility_specialcase} illustrates
the performance gap between the MED mechanism and the optimal mechanism
in the average case. We use the \emph{relative difference} metric
which is defined as the ratio of the SC platform's utility difference
achieved from the OPT mechanism and the MED mechanism over the SC
platform's utility achieved from the OPT mechanism, i.e., $\frac{\hat{u}{}_{m}(\mathsf{M}_{\mathrm{OPT}}(\boldsymbol{L}))-\hat{u}{}_{m}(\mathsf{M}_{\mathrm{MED}}(\boldsymbol{L}))}{\hat{u}{}_{m}(\mathsf{M}_{\mathrm{OPT}}(\boldsymbol{L}))}$.
When the radio environment gets worse (a larger $\alpha$), we can
see the decreasing performance of the MED mechanism compared to the
optimal mechanism. However, if more workers participate, the performance
of the MED mechanism is closer to the optimal results. This is mainly
due to the fact that the generated workers' locations follows a similar
symmetric uniform distribution.

\section{Conclusion\label{sec:Conclusion}}

In this paper, we have proposed a wireless powered spatial crowdsourcing
framework composed of two phases. We have proven that the proposed
Stackelberg game based incentive mechanism can help the SC platform
efficiently allocate the tasks and the wireless charging power in
the task allocation phase. For the deployment of the mobile BS in
the data crowdsourcing phase, we have adopted the median mechanism
as the strategyproof mobile BS deployment mechanism. Besides avoiding
the dishonest worker's manipulation, the proposed SC framework can
efficiently allocate tasks and charging power which is shown by experimental
results. 

\section*{{\small{}Acknowledgment}}

{\footnotesize{}This work was supported in part by A{*}STAR-NTU-SUTD
Joint Research Grant Call on Artificial Intelligence for the Future
of Manufacturing RGANS1906, WASP/NTU M4082187 (4080), Singapore MOE
Tier 1 under Grant 2017-T1-002-007 RG122/17, MOE Tier 2 under Grant
MOE2014-T2-2-015 ARC4/15, Singapore NRF2015-NRF-ISF001-2277, Singapore
EMA Energy Resilience under Grant NRF2017EWT-EP003-041, Nanyang Technological
University (NTU) Startup Grant M4082311.020, Singapore Ministry of
Education Academic Research Fund Tier 1 RG128/18, NTU-WASP Joint Project
M4082443.020, Canada NSERC Discovery grant RGPIN-2019-06375 and National
Research Foundation of Korea (NRF) Grant funded by the Korean Government
under Grant 2014R1A5A1011478.}{\footnotesize\par}

\bibliographystyle{IEEEtran}
\bibliography{IEEEabrv,PC4_location_mechanism_sc}

\begin{thebibliography}{10}
\providecommand{\url}[1]{#1}
\csname url@samestyle\endcsname
\providecommand{\newblock}{\relax}
\providecommand{\bibinfo}[2]{#2}
\providecommand{\BIBentrySTDinterwordspacing}{\spaceskip=0pt\relax}
\providecommand{\BIBentryALTinterwordstretchfactor}{4}
\providecommand{\BIBentryALTinterwordspacing}{\spaceskip=\fontdimen2\font plus
\BIBentryALTinterwordstretchfactor\fontdimen3\font minus
  \fontdimen4\font\relax}
\providecommand{\BIBforeignlanguage}[2]{{%
\expandafter\ifx\csname l@#1\endcsname\relax
\typeout{** WARNING: IEEEtran.bst: No hyphenation pattern has been}%
\typeout{** loaded for the language `#1'. Using the pattern for}%
\typeout{** the default language instead.}%
\else
\language=\csname l@#1\endcsname
\fi
#2}}
\providecommand{\BIBdecl}{\relax}
\BIBdecl

\bibitem{Kazemi2012}
L.~Kazemi and C.~Shahabi, ``Geocrowd: enabling query answering with spatial
  crowdsourcing,'' in \emph{Proceedings of the 20th international conference on
  advances in geographic information systems}.\hskip 1em plus 0.5em minus
  0.4em\relax ACM, 2012, pp. 189--198.

\bibitem{Zhao2016}
Y.~Zhao and Q.~Han, ``Spatial crowdsourcing: current state and future
  directions,'' \emph{IEEE Communications Magazine}, vol.~54, no.~7, pp.
  102--107, 2016.

\bibitem{Guo2018}
B.~Guo, Y.~Liu, L.~Wang, V.~O. Li, C.~Jacqueline, and Z.~Yu, ``Task allocation
  in spatial crowdsourcing: Current state and future directions,'' \emph{IEEE
  Internet of Things Journal}, 2018.

\bibitem{Peng2010}
Y.~Peng, Z.~Li, W.~Zhang, and D.~Qiao, ``Prolonging sensor network lifetime
  through wireless charging,'' in \emph{2010 31st IEEE Real-Time Systems
  Symposium}, Nov 2010, pp. 129--139.

\bibitem{Li2018}
K.~Li, W.~Ni, L.~Duan, M.~Abolhasan, and J.~Niu, ``Wireless power transfer and
  data collection in wireless sensor networks,'' \emph{IEEE Transactions on
  Vehicular Technology}, vol.~67, no.~3, pp. 2686--2697, March 2018.

\bibitem{Bi2015}
S.~Bi, C.~K. Ho, and R.~Zhang, ``Wireless powered communication: Opportunities
  and challenges,'' \emph{IEEE Communications Magazine}, vol.~53, no.~4, pp.
  117--125, 2015.

\bibitem{Restuccia2016}
F.~Restuccia, S.~K. Das, and J.~Payton, ``Incentive mechanisms for
  participatory sensing: Survey and research challenges,'' \emph{ACM
  Transactions on Sensor Networks (TOSN)}, vol.~12, no.~2, p.~13, 2016.

\bibitem{Yang2016}
D.~Yang, G.~Xue, X.~Fang, and J.~Tang, ``Incentive mechanisms for crowdsensing:
  Crowdsourcing with smartphones,'' \emph{IEEE/ACM Trans. Netw.}, vol.~24,
  no.~3, pp. 1732--1744, Jun. 2016.

\bibitem{Moulin1980}
H.~Moulin, ``On strategy-proofness and single peakedness,'' \emph{Public
  Choice}, vol.~35, no.~4, pp. 437--455, 1980.

\bibitem{Dieter2005}
W.~R. Dieter, S.~Datta, and W.~K. Kai, ``Power reduction by varying sampling
  rate,'' in \emph{Proceedings of the 2005 international symposium on Low power
  electronics and design}.\hskip 1em plus 0.5em minus 0.4em\relax ACM, 2005,
  pp. 227--232.

\bibitem{Zhou2013}
X.~Zhou, R.~Zhang, and C.~K. Ho, ``Wireless information and power transfer:
  Architecture design and rate-energy tradeoff,'' \emph{IEEE Transactions on
  Communications}, vol.~61, no.~11, pp. 4754--4767, November 2013.

\bibitem{Kahneman2013}
D.~Kahneman and A.~Tversky, ``Prospect theory: An analysis of decision under
  risk,'' in \emph{Handbook of the fundamentals of financial decision making:
  Part I}.\hskip 1em plus 0.5em minus 0.4em\relax World Scientific, 2013, pp.
  99--127.

\bibitem{Fudenberg1991}
D.~Fudenberg and J.~Tirole, \emph{Game Theory}.\hskip 1em plus 0.5em minus
  0.4em\relax Cambridge, MA, USA: MIT Press, 1991.

\bibitem{Rosen1965}
\BIBentryALTinterwordspacing
J.~B. Rosen, ``Existence and uniqueness of equilibrium points for concave
  n-person games,'' \emph{Econometrica}, vol.~33, no.~3, pp. 520--534, 1965.
  [Online]. Available: \url{http://www.jstor.org/stable/1911749}
\BIBentrySTDinterwordspacing

\bibitem{Han2012}
Z.~Han, D.~Niyato, W.~Saad, T.~Baar, and A.~Hjrungnes, \emph{Game Theory in
  Wireless and Communication Networks: Theory, Models, and Applications},
  1st~ed.\hskip 1em plus 0.5em minus 0.4em\relax New York, NY, USA: Cambridge
  University Press, 2012.

\bibitem{Barbera1993}
S.~Barber{\`a}, F.~Gul, and E.~Stacchetti, ``Generalized median voter schemes
  and committees,'' \emph{Journal of Economic Theory}, vol.~61, no.~2, pp.
  262--289, 1993.

\bibitem{Feldman2013}
M.~Feldman and Y.~Wilf, ``Strategyproof facility location and the least squares
  objective,'' in \emph{Proceedings of the Fourteenth ACM Conference on
  Electronic Commerce}, ser. EC '13.\hskip 1em plus 0.5em minus 0.4em\relax New
  York, NY, USA: ACM, 2013, pp. 873--890.

\bibitem{Jameson2014}
G.~J.~O. Jameson, ``Some inequalities for $(a+b)^p$ and $(a+b)^p+(a-b)^p$,''
  \emph{The Mathematical Gazette}, vol.~98, no. 541, pp. 96--103, 2014.

\end{thebibliography}

\end{document}